\documentclass[]{aa}
\usepackage{natbib}
\usepackage{epsfig}
\def\ut#1{\mathop{\vtop{\ialign{##\crcr
     $\hfil\displaystyle{#1}\hfil$\crcr\noalign
     {\kern1pt\nointerlineskip}\hbox{$\hfil\sim\hfil$}\crcr
     \noalign{\kern1pt}}}}}

\def\undersymbol#1#2{\mathop{\vtop{\ialign{##\crcr
     $\hfil\displaystyle{#2}\hfil$\crcr\noalign
     {\kern1pt\nointerlineskip}\hbox{$\hfil#1\hfil$}\crcr
     \noalign{\kern1pt}}}}}

\begin{document}

\title{Monte Carlo analysis of MEGA microlensing events towards M31}

\author{G. Ingrosso\inst{1},
        S. Calchi Novati\inst{2},
        F. De Paolis\inst{1},
        Ph. Jetzer\inst{2},
        A.A. Nucita\inst{1},
        F. Strafella\inst{1}}
%\offprints{xx, \\
 %   \email{jetzer@physik.unizh.ch}}
\institute{Dipartimento di Fisica,
           Universit\`{a} di Lecce and INFN, Sezione di Lecce,
           CP 193, I-73100 Lecce, Italy \and
           Institute for Theoretical Physics,
           University of  Z\"{u}rich, Winterthurerstrasse 190,
           CH-8057 Z\"{u}rich, Switzerland}
\date{Received  / Accepted}
\authorrunning{Ingrosso et al.}
\titlerunning{Monte Carlo analysis of the MEGA microlensing events
towards M31}
%---------------------------------------------------------------
\abstract{
We perform an analytical study and a Monte Carlo (MC) analysis of the main
features for microlensing events in pixel lensing observations towards
M31. Our main aim is to investigate the lens nature and location
of the 14 candidate events found by the MEGA collaboration.
Assuming a reference model for the mass distribution in M31 and the
standard model for our galaxy,
we estimate the MACHO-to-self lensing probability
and the event time duration towards M31.
Reproducing the MEGA observing conditions,
as a result we get the MC event number density distribution as
a function of the event full-width half-maximum
duration $t_{1/2}$ and the magnitude at maximum
$R_{\mathrm {max}}$.
For a MACHO mass of
$0.5~M_{\odot}$ we find typical values of $t_{1/2} \simeq 20$ day
and $R_{\mathrm {max}} \simeq 22$,
for both MACHO-lensing and self-lensing events occurring
beyond about 10 arcminutes from the M31
center.
A comparison of the observed features
($t_{1/2}$ and $R_{\mathrm {max}}$) with our MC results
shows that for a MACHO mass $>0.1~M_{\odot}$
the four innermost MEGA events are most likely self-lensing events,
whereas the six outermost events must be genuine MACHO-lensing events.

}

\maketitle

%-----------------------------------------------------------------
\section{Introduction}
%-----------------------------------------------------------------
Gravitational microlensing has become since about a decade a robust tool
for analyzing the galactic structure and for gaining information
about the dark mass component in our galaxy \citep{Alcock93,Aubourg93}.
Several hundreds of microlensing events have been detected so far
towards the galactic bulge, the spiral arms, and the Magellanic Clouds
\citep{Alcock00,jetzer}.

Recently, pixel lensing observations towards M31 and even M87 have
been undertaken and some microlensing events have been found.
The MEGA collaboration \citep{Mega04}
has reported the detection of 14 candidate events towards M31 by
using Isaac Newton Telescope (INT) data.
Two of these events have been previously reported
by the POINT-AGAPE collaboration \citep{Paulin03},
which recently has presented a high-threshold
analysis of the full 3 years data set
\citep{Calchi05}. This analysis shows that the observed
signal is much larger than expect from self-lensing alone and that
some fraction of the halo mass must be in form
of MACHOs.
Other collaborations have also reported preliminary
results for pixel lensing towards M31
\citep{Calchi02,Calchi03,Uglesich04,Wecapp03,joshi04}.

In this paper we consider in particular the MEGA results which provide
the largest sample of microlensing candidates.
A map of the MEGA events is reported by de Jong et al.
(\citeyear{Mega04}). A preliminary analysis by the MEGA collaboration
indicates that the events located in the outer part of M31 are
consistent with being due to halo lens objects, whereas the
innermost ones are most likely due to self-lensing.
The aim of this paper is to perform a more complete analysis by
using a Monte Carlo (MC) program with the purpose of using all
available information by the MEGA collaboration, which are summarized
in Tab. \ref{Mega14post12df},
to characterize the nature of the lenses.

%%%%%%%%%%%%%%%%%%%%%%%%%%%%%%

\begin{table}
%\centering
\caption{For the 14 MEGA events we give the position,
the full-width half-maximum duration $t_{1/2}$
and the magnitude at maximum
$R_{\mathrm max}$. The coordinate system we adopt
has origin in the M31 center and the
X axis is oriented along the M31 disk major
axis (see also Fig. \ref{uth}).}
\medskip
\begin{tabular}{|c|c|c|c|c|}
\hline
MEGA & X & Y      & $t_{1/2}^{~~~\mathrm obs}$  & $R_{max}^{~~~\mathrm obs}$\\
     & arcmin & arcmin & day        & magn         \\
\hline
1&   -4.367&  -2.814&    4.20  $\pm$  4.30&    22.2 $\pm$    1.1\\
2&   -4.478&  -3.065&    4.60  $\pm$  0.60&    21.6 $\pm$    0.3\\
3&   -7.379&  -1.659&    2.60  $\pm$  2.20&    21.8 $\pm$    1.2\\
4&  -10.219&   3.420&   29.10  $\pm$  1.00&    22.8 $\pm$    0.2\\
5&  -19.989& -13.955&    9.40  $\pm$  4.10&    22.9 $\pm$    0.8\\
6&  -21.564& -13.169&   22.90  $\pm$  0.70&    22.6 $\pm$    0.2\\
7&  -21.163&  -6.230&   21.60  $\pm$  0.70&    19.3 $\pm$    0.2\\
8&  -21.650&   7.670&   27.40  $\pm$  0.90&    22.7 $\pm$    0.2\\
9&  -33.834&  -2.251&    3.80  $\pm$  1.60&    21.8 $\pm$    0.8\\
10&  -3.933& -13.846&   46.80  $\pm$  4.40&    22.2 $\pm$    0.3\\
11&  19.192& -11.833&    2.00  $\pm$  0.30&    20.5 $\pm$    0.2\\
12&  29.781&  -5.033&  131.00  $\pm$  9.40&    23.2 $\pm$    0.3\\
13&  22.072& -22.022&   22.80  $\pm$  3.80&    23.3 $\pm$    0.3\\
14&  19.348& -29.560&   28.10  $\pm$  1.40&    22.5 $\pm$    0.2\\
\hline
\end{tabular}
%\end{center}
\label{Mega14post12df}
\end{table}

The paper is organized as follows: in Section 2 we present the
calculation of the microlensing rate towards M31 and
we remind the pixel lensing basics.
In Section 3 we give some details on the adopted M31 and Milky
Way (MW) mass distribution
models as well as the stellar and mass functions.
The analytic and MC results and a comparison
between the two are given in Sections 4-6 and
conclusions are presented in Section 7.

%------------------------------------------------------------------------
\section{Microlensing rate}

%------------------------------------------------------------------------

The differential number of expected microlensing events
is  \citep{djm,Griest}
\begin{equation}
dN_{\mathrm{ev}}=N_{*}t_{\mathrm{obs}}d\Gamma,
\end{equation}
where $N_{*}$ is the total number of monitored stars during the
observation time $t_{\mathrm{obs}}$. The differential
rate $d\Gamma$ at which a single star is microlensed by a compact object
is given by
\begin{equation}
d\Gamma= \frac{n_{\mathrm l}(\vec{x},\mu) d\mu
f(\vec v_{\mathrm{l \perp}}) d^{2} \vec v_{\mathrm l \perp} d^{3}x }{dt},
\end{equation}
where the numerator on the right hand side is the number of lenses with
transverse velocity
in $ d^{2} \vec v_{\mathrm l \perp}
= v_{\mathrm l \perp }~ dv_{\mathrm l \perp }~d\beta$ around
$ \vec v_{\mathrm{l \perp}}$, located in a
volume element $d^{3}x=dx~ dy~ dz$ centered at the position $\vec{x}$ of the
microlensing tube. Here $\vec v_{\mathrm l \perp}$ is the component
(in the rest frame of the Galaxy) of the lens velocity
orthogonal to the line of sight to the source star,
$n_{\mathrm l}(\vec{x},\mu)$ the lens number density (per unit of
volume and mass) and
$f(\vec{v}_{\mathrm{l \perp}})$ the lens tranverse velocity distribution.
$\mu$ is the lens mass in solar units.

In evaluating $d\Gamma$
we must take into account that the source
stars are not at rest but have a transverse velocity
$\vec{v}_{\mathrm s \perp}$ as well, which can be splitted into a random
component and a component which describes the ordered rotation (if present)
of the
galactic component considered, namely
$\vec{v}_{\mathrm{s \perp}} = \vec{v}_{\mathrm{s \perp,ran}} +
\vec{v}_{\mathrm{s \perp,rot}}$.
Using a Maxwellian distribution with 2-dimensional
velocity dispersion $\sigma_{\mathrm{s}}$
to describe the random velocity component,
we get for the transverse velocity distribution
$g(\vec{v}_{\mathrm{s \perp}})$
\begin{equation}
\label{vs_distribution}
g(\vec{v}_{\mathrm{s \perp}})
v_{\mathrm{s \perp}} dv_{\mathrm{s \perp}} d\varphi =
\frac{1}{\pi \sigma_{\mathrm{s}}^{2}}
e^{-\frac{(\vec v_{\mathrm{s \perp}}-\vec v_{\mathrm{s \perp,rot}})^{2}}
{\sigma_{\mathrm{s}}^{2}}}
v_{\mathrm{s \perp}} dv_{\mathrm{s \perp}} d\varphi,
\label{gg}
\end{equation}
where $\varphi$ is the angle between $\vec v_{\mathrm{s  \perp}}$ and
$\vec v_{\mathrm{s \perp,rot}}$.

We take also into account the transverse
rotation velocity
of the observer, $\vec{v}_{\mathrm \sun \perp,rot}$
(here we consider the observer
co-moving with the Sun) by considering that the microlensing tube is moving
with a transverse velocity
\begin{equation}
\label{tube-velocity}
\vec{v}_{\mathrm{t \perp}}=
\left(1-\frac{D_{\mathrm{ol}}}{D_{\mathrm{os}}}\right)
\vec{v}_{\mathrm \sun \perp,rot}+
\frac{D_{\mathrm{ol}}}{D_{\mathrm{os}}}
\vec{v}_{\mathrm{s \perp}}~,
\end{equation}
where $D_{\mathrm {os}}$ and $D_{\mathrm{ol}}$ are the source and lens
distances from the observer, respectively.

As for the source stars
we split the lens transverse velocity into a random component
and an ordered rotation component. Moreover, by taking into account that
the microlensing tube moves with velocity $\vec{v}_{\mathrm{t \perp}}$,
it follows that
$\vec v_{\mathrm l \perp} = \vec v_{\mathrm l \perp,ran} +
\vec v_{\mathrm l \perp,rot}-\vec v_{\mathrm t \perp}$.
Accordingly, assuming for $\vec v_{\mathrm l \perp,ran}$ a Maxwellian
distribution
with 2-dimensional velocity dispersion $\sigma_l$, we obtain
using also the definition for $g(\vec{v}_{\mathrm{s \perp}})$
given in eq.(\ref{gg})
\begin{eqnarray}
f(\vec v_{\mathrm{l \perp}})
v_{\mathrm l \perp } dv_{\mathrm l \perp }~d\beta =
\frac{1}{(\pi \sigma_{\mathrm s } \sigma_{\mathrm l})^2}
\int_0^{2 \pi} d\varphi  \times \nonumber \\
\int_0^{\infty}
e^{-\frac{(\vec v_{\mathrm s \perp} -\vec v_{\mathrm s \perp,rot})^{2}}
{\sigma_{\mathrm s }^{2}}}
v_{\mathrm s \perp} dv_{\mathrm s \perp}  ~
e^{-\frac{(\vec v_{\mathrm l \perp} - \vec w_{\perp})^2}
{\sigma_{\mathrm l }^2}}
v_{\mathrm l \perp } dv_{\mathrm l \perp }~d\beta~,
\label{2020}
\end{eqnarray}
where $\vec w_{\perp}= \vec v_{\mathrm l \perp,rot}-\vec
v_{\mathrm t \perp}$ and
$\beta$ is the angle between $\vec v_{\mathrm{l \perp}}$ and
$\vec w_{\perp}$.

We can then write the volume element, $d^{3}x$, as
\begin{eqnarray}
d^{3}x&=&(\vec{v}_{\mathrm{l \perp}}\cdot\hat{n})~dt~dS= v_{\mathrm{l \perp}}
\cos{\theta}~dt~dl~dD_{\mathrm{ol}} \\
&=& v_{\mathrm{l \perp}}\cos{\theta}~dt~R_{\mathrm{E}}~du_{\mathrm{th}}~
d{\alpha}~dD_{\mathrm{ol}}\nonumber,
\end{eqnarray}
where $\theta$ is the angle  between
$\vec{v}_{\mathrm{l \perp}}$ and the normal, $\hat{n}$,
to the lateral superficial element, $dS=dldD_{\mathrm{ol}}$,
of the microlensing tube, with
$dl=R_{\mathrm{E}}du_{\mathrm{th}}d{\alpha}$
being the cylindrical segment of the tube
($u_{\mathrm{th}}$ is the threshold value for the impact parameter).
Note that, if $\alpha$ is taken to be
the angle between $\hat{n}$ and $\vec w_{\perp}$, it
follows that $\theta = \alpha + \beta $,
so that for a constant value of $\beta$, $d\alpha=d\theta$.
Therefore, the microlensing differential event rate becomes
\begin{equation}
d\Gamma= n_{\mathrm l}(\vec x, \mu) d\mu
f(\vec{v}_{\mathrm{l \perp}}) v_{\mathrm{l \perp}}^{2}
dv_{\mathrm{l \perp}} d{\beta}
\cos{\theta} R_{\mathrm{E}} du_{\mathrm{th}}
d{\alpha} dD_{\mathrm{ol}} ~.
\label{1991}
\end{equation}

We assume, as usual, that
the mass distribution of the lenses is independent
of their position in M31 or in the Galaxy
({\it factorization hypothesis}).
So, the lens number density
$n_{\mathrm l} (\vec x, \mu)$ can be written as \citep{jms}
\begin{equation}
n_{\mathrm l}(\vec x, \mu)=
\frac{ \rho_{\mathrm l}(\vec x)} {\rho_0} ~ \psi_0(\mu)~,
\label{psi0}
\end{equation}
where  $\rho_0$ is the local mass density in the Galaxy
or the central density in M31,
$\psi_0(\mu)$ the corresponding
lens number density per unit of mass and
the normalization is
\begin{equation}
\int_{\mu_{\mathrm{min}}}^{\mu_{\mathrm{up}}}
\psi_0(\mu) ~ \mu ~d\mu~ = \frac{\rho_0}{M_{\odot}}~.
\label{norm}
\end{equation}
Here $\mu_{\mathrm {min}}$ and $\mu_{\mathrm {up}}$ are the lower and the
upper limits for the lens masses (see Subsect. 3.3).
%%%%%%%%%%%%%%%%%%%%%%%%%%%%%%%%%%%%%%%%%%%%%%%%%%%%%%%%
\begin{figure}[htbp]
\vspace{7.0cm} \includegraphics{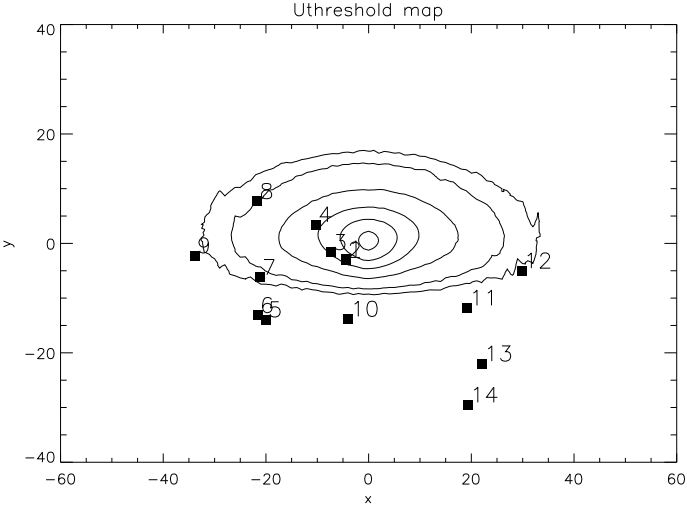}
\caption{
The $\langle u_{\mathrm T} (x,y) \rangle_{\phi_{\mathrm RG}}$ contour
plot towards the M31 galaxy is shown with the position in the sky plane
of the 14 MEGA candidate microlensing events.
From the inner to the outer part of the figure
$\langle u_{\mathrm T} (x,y) \rangle_{\phi_{\mathrm RG}}$ increases
with lines referring to the values 0.02, 0.04, 0.06, 0.08 and 0.10,
respectively. Outside the 0.10 contour line
$\langle u_{\mathrm T} (x,y) \rangle_{\phi_{\mathrm RG}}$
is almost constant.
Note that the positions of the events 1 and 2 are almost
on top of each other. The same is true as well for all other figures.
}
\label{uth}
\end{figure}
%%%%%%%%%%%%%%%%%%%%%%%%%%%%%%%%%%%%%%%%%%%%%%%%%%%%%%%%
Accordingly, the microlensing event rate is given by:
\begin{eqnarray}
\label{gamma1}
 \Gamma ( D_{\mathrm{os}}) =
\sqrt{\frac{4GM_{\sun}}{c^{2}}}
\int_{\mu_{\mathrm{min}}}^{\mu_{\mathrm{up}}} d\mu ~\mu^{1/2}~ \psi_0(\mu)
\int_0^{u_{\mathrm T}} du_{\mathrm th}
\times \nonumber \\
\int_{0}^{D_{\mathrm{os}}} dD_{\mathrm{ol}}
\sqrt{\frac{D_{\mathrm{ol}}(D_{\mathrm{os}}-D_{\mathrm{ol}})}
{D_{\mathrm{os}}}} ~ \frac{\rho_{\mathrm l} (D_{\mathrm{ol}})}{\rho_0}
~\times
\nonumber \\
\int_0^{2 \pi} d\beta
\int_0^{\infty} dv_{\mathrm l \perp }
f(\vec v_{\mathrm{l \perp}})
v^2_{\mathrm l \perp }
\int_{-\pi/2}^{+\pi/2} \cos \theta d\theta~,
\end{eqnarray}
where the integration on $\theta$ is performed between
$-\pi/2$ and $+\pi/2$, since only lenses entering the microlensing tube
are considered.
After integrations on $\theta$ and $\beta$ we finally get
\begin{eqnarray}
\label{gamma}
\Gamma ( D_{\mathrm{os}}) =
2 \sigma_{\mathrm l} u_{\mathrm T}
\sqrt{\frac{4GM_{\sun}}{c^{2}}}
\int_{\mu_{\mathrm{min}}}^{\mu_{\mathrm{up}}} d\mu ~ \mu^{1/2}~ \psi_0(\mu)
\times \nonumber \\
\int_{0}^{D_{\mathrm{os}}} dD_{\mathrm{ol}}
\sqrt{\frac{D_{\mathrm{ol}}(D_{\mathrm{os}}-D_{\mathrm{ol}})}
{D_{\mathrm{os}}}} \frac{\rho_{\mathrm{l}}(D_{\mathrm{ol}})}{\rho_0}
\int_0^{\infty} P(z) dz~,
\label{gamma11}
\end{eqnarray}
where $ z = v_{\mathrm l \perp}/\sigma_{\mathrm l}$
and the function $P(z)$ is given by
\begin{eqnarray}
P(z) = \frac{2 e^{-a^2} }{\pi}
\int_0^{2 \pi} d\varphi
\int_0^{\infty}
 ye^{-(y^2 - 2ay\cos\varphi+\eta^2)} dy
\times \nonumber \\
z^2 e^{-z^2} I_0(2\eta z)~.
\label{provt}
\end{eqnarray}
Here
$a(D_{\mathrm{os}}) = v_{\mathrm s \perp,rot}/\sigma_{\mathrm s }$,
$y=v_{\mathrm s \perp}/\sigma_{\mathrm s }$,
$\eta( D_{\mathrm{os}}, y, \varphi)
 = w_{\mathrm \perp}/\sigma_{\mathrm l }$  and
$I_0(2\eta z)$ is the zero-order modified Bessel function of the argument
$2\eta z$.

Note that $\chi(z) = z^{-1} P(z)$ is the dimensionless form of eq. (\ref{2020})
and that it is properly normalized
%($\int_0^{\infty} \chi(z) dz =1$)
as it can be easily verified
by using twice the relation
$\int_0^{\infty} dt~ exp(-t^2)~ t~ I_0(2qt) = exp(q^2)/2$.

To take into account the source distribution in the M31 bulge and disk,
eq. (\ref{gamma}) has to be integrated not only over the distance
of the lenses but also over the distance of the sources.
Accordingly, the microlensing rate becomes
\begin{equation}
\label{microlensing-rate}
\Gamma (x,y)=
\frac{\int_0^{\infty} \rho_s(D_{\mathrm {os}}) \Gamma(D_{\mathrm {os}})
d D_{\mathrm {os}} }
{\int_0^{\infty} \rho_s(D_{\mathrm {os}}) d D_{\mathrm {os}} }~,
\end{equation}
where $x$ and $y$ are coordinates in the plane orthogonal to line
of sight,  $\rho_s$ is the source mass density (which is the
sum of the sources in the M31 bulge and disk).

Moreover,  we compute the average Einstein time (which depends
on the line of sight position given by the coordinates $x$ and $y$) as
\begin{equation}
<t_E> =
\frac{\int_0^{\Gamma}
t_E d\Gamma }{\Gamma}~.
\end{equation}

\subsection{Pixel lensing basics}

Pixel lensing technique is based on the observation of
the flux variations of every element (pixel) of an image
\citep{Ansari97}.
Looking towards M31 a large number
of stars contribute at the same time to the flux received by each pixel
so that only highly magnified events can be detected.

To be detectable a microlensing event must give rise to
a substantial flux variation with respect to the background
$N_{\mathrm {bl}} = N_{\mathrm {gal}} + N_{\mathrm {sky}}$,
which is the sum of the M31 surface brightness
and the sky contribution.
The excess photon count per pixel due to an ongoing microlensing event is
\begin{equation}
\Delta N_{\mathrm {pix}} =
       N_{\mathrm {bl }}
      [A_{\mathrm {pix}}-1] =
       f_{\mathrm {see}}
       N_{\mathrm s} [A(t)-1]
\end{equation}
where $N_{\mathrm s}$ is the source photon count in the absence of lensing,
$A(t)$ is the source magnification factor due to lensing
(see e.g. \citealt{Griest})
and $f_{\mathrm {see}}$ the fraction of the seeing disk contained in a pixel.
Therefore, the expected number of photons in a pixel will be
$N_{\mathrm {pix}} = N_{\mathrm {bl}} + \Delta N_{\mathrm {pix}}$.

Of course, a pixel lensing event is detectable if the excess pixel photon
count is greater than the threshold pixel noise $\sigma_{\mathrm T}$.
Accordingly, by requiring the signal to be at least $3 \sigma_{\mathrm T}$
level above the baseline count,
one obtains a threshold value for the amplification
\citep{Kerins01}
\begin{equation}
A_{\mathrm T} = 1+\frac{3 \sigma_{\mathrm T}}
{f_{\mathrm {see}} N_{\mathrm s}}~,
\label{AT}
\end{equation}
which corresponds to a threshold value $u_{\mathrm T}$ for the impact
parameter,
via the well known relation between the lens impact parameter
and the amplification factor.

One can estimate $\sigma_{\mathrm T}$ as the
maximum between the statistical error $\propto \sqrt{N_{\mathrm {bl}}}$
and $\simeq 3 \times 10^{-3} ~ N_{\mathrm {bl}}$ that is
determined by the pixel flux stability.
Accordingly, $u_{\mathrm T}$ depends on both the line of sight to M31
and the source magnitude $M$.

Hence, by averaging on the source luminosity function $\phi (M)$,
we can evaluate the average threshold impact parameter for any
direction towards the M31 galaxy, so that we get
\begin{equation}
\langle u_{\mathrm T} (x,y) \rangle_{\phi}
= \frac{\int  u_{\mathrm T} (x,y;M)~ \phi (M) dM}
  {\int  \phi(M) dM}~,
\label{utmediosuphi}
\end{equation}
where the coordinates $x$ and $y$ span the sky plane towards M31.

By using the threshold impact parameter defined in eq. (\ref{utmediosuphi}),
one obtains the pixel lensing rate as follows
\citep{Kerins01,Kerins03,dinz}
\begin{equation}
\Gamma_p (x,y) = \langle u_T(x,y) \rangle_{\phi} ~\Gamma (x,y)~.
\label{gammapix}
\end{equation}

%%%%%%%%%%%%%%%%%%%%%%%%%%%%%%%%%%%%%%%%%%%%%%%%%%%%%%%%
\begin{figure}[htbp]
\vspace{7.0cm} \includegraphics{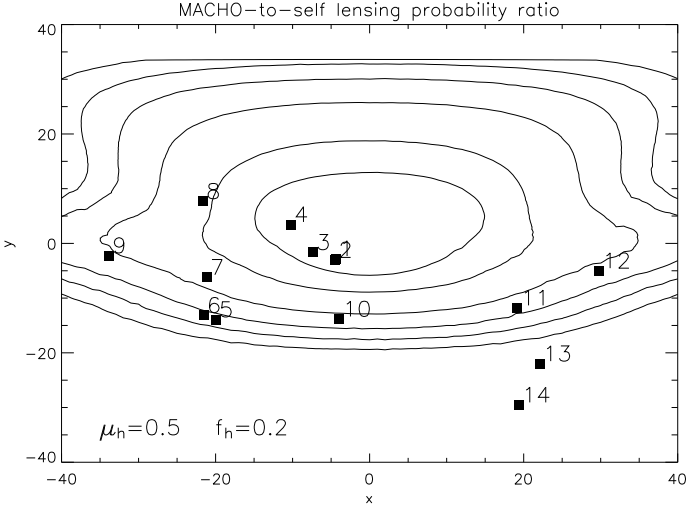}
\caption{MACHO-to-self lensing probability ratio
$(P_{\mathrm h }/P_{\mathrm s})_{\mathrm {An}}$ map
projected onto the sky plane.
Here and in the following figures we
assume a fraction $f_{\rm h} = 20\%$ of dark matter
and a MACHO mass $\mu_{\mathrm h} = 0.5$,
both in M31 and MW halos.
From the inner to the outer part
$(P_{\mathrm h}/P_{\mathrm s})_{\mathrm {An}}$
increases with lines referring to the values 0.5, 1, 2, 3, 4 and 5,
respectively.}
\label{prodar}
\end{figure}
%%%%%%%%%%%%%%%%%%%%%%%%%%%%%%%%%%%%%%%%%%%%%%%%%%%%%%%%
%%%%%%%%%%%%%%%%%%%%%%%%%%%%%%%%%%%%%%%%%%%%%%%%%%%%%%%%
\begin{figure}[htbp]
\vspace{7.0cm} \includegraphics{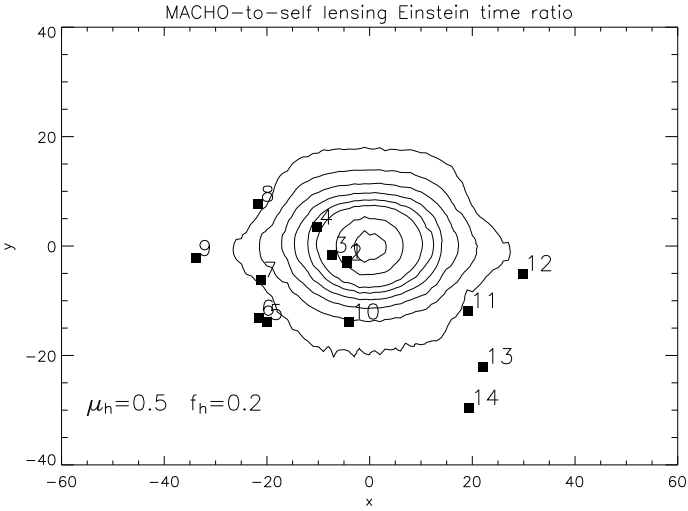}
\caption{The MACHO-to-self lensing Einstein time ratio
$<t_{E~\mathrm h}>/<t_{E~\mathrm s}>$ map is given
for $\mu_{\mathrm h} = 0.5$. Going from the inner to the outer part
the ratio decreases with lines corresponding to values in between 2 to 1.25
with a step of 0.25.}
\label{timeratio}
\end{figure}
%%%%%%%%%%%%%%%%%%%%%%%%%%%%%%%%%%%%%%%%%%%%%%%%%%%%%%%%%%%%

\section{Modeling}

\subsection{M31 and Galaxy mass distribution models}

The M31 disk, bulge and halo mass distributions
are described adopting the reference
model discussed in \cite{Kerins04}.
This model, providing remarkable good fits to the M31 surface brightness
and rotation curve profiles, can be considered as an acceptable
model for the mass distribution in the M31 galaxy.
Accordingly,
the mass density of the M31 disk stars is described by
a sech-squared profile
\begin{equation}
\rho_D (R,z) = \rho_D(0) \exp(-R/h)~ {\rm sech}^2(z/H),
\label{discoprofile}
\end{equation}
where $H=0.3$ kpc, $h=6.4$ kpc and
$\rho_D(0)= 0.35 \times 10^9 ~M_{\odot}$
pc$^{-3}$ are, respectively,
the scale height and scale lengths of the disk and
the disk central mass density.
$R$ is the distance on the M31 disk plane
(described by the coordinates $x$ and $y$) and $z$ is the distance
from it.
The M31 disk is assumed to be inclined by an angle
$i=77^{0}$  and the azimuthal angle relative to the near minor
axis is $\phi = -38.6^{0}$.

The M31 bulge is parameterized by a flattened power law of the form
\begin{equation}
\rho_B(R,z) = \rho_B(0) \left[ 1+ \left( \frac {R}{\tilde
a}\right)^2 +q^{-2}
\left( \frac {z}{\tilde a}\right)^2\right]^{-s/2}~,
\end{equation}
where $\rho_B(0) \simeq 4.5 \times 10^9~M_{\odot}$ kpc$^{-3}$,
$ q \simeq 0.6$ is the ratio of the minor to major axis,
$\tilde a \simeq $ 1 kpc and $s \simeq 3.8$.
Both the M31 disk and bulge are truncated at a distance $R = 40$ kpc.

We remark that the twisting of the optical isophotes in the inner
M31 regions indicates that the bulge major axis
is offset by $\simeq 15^0$ from the disk major axis \citep{sb}.
The consideration of this effect by Kerins et al. \citeyear{Kerins05}
leads to the evaluation of pixel lensing rates that show spatial distributions
tilted of the same amount inside 5 arcmin from the M31 center.
The twisting effect vanishes at larger distances
due to the increasing contribution to microlensing by M31 disk and halo.
Clearly, our results in Figs. \ref{uth}-\ref{timeratio}
do not show the above mentioned effect
since here we are considering only a flattened bulge without twist.
However, we expect that the consideration of the isophote twisting
does not substantially modify our results about both
MACHO-to-self lensing probability and event time scale ratios
(given in Tabs. \ref{probrate}-\ref{probratemc}), particularly
for events at large distance from the M31 center.

The dark matter in the M31 halo is assumed to follow an isothermal
profile
\begin{equation}
\rho_H(r) = \rho_H(0) \frac{a^2}{a^2+r^2}~,
\label{haloprofile}
\end{equation}
with core radius $a=4$ kpc and central dark matter
density $\rho_H(0)= 6.5 \times 10^7 ~M_{\odot}$ kpc$^{-3}$.
The M31 halo is truncated at 100 kpc with
asymptotic rotational velocity $v_{rot} \simeq 235$ km s$^{-1}$.

We do not consider other dark matter distribution models,
as King-Michie \citep{Binney} or NFW \citep{NFW} suggested by N body
simulations.
The effect of using such models, which have a more concentred dark mass
distribution, is both to decrease the spatial distribution of MACHOs at large
distance from the M31 center, where the rotation curve is poorly
determined, and to increase it in the innermost region, where the MACHO
contribution is relatively unimportant with respect to that of bulge and disk.
Regarding the former aspect, we also note that
in the MEGA experiment the typical MACHO lens distance (about 20 kpc)
is too small to appreciate the effect of this choice.
Overall, the current data do not allow one to perform a
meaningfully fine tuning of the dark matter parameters.

As usual, the mass density profile for the MW disk is
described with a double exponential profile
\begin{equation}
\rho_D (R,z) = \rho_D(R_0) \exp(-(R-R_0)/h)~\exp(-|z|/H)~,
\end{equation}
with Earth position from the Galactic center at $R_0 \simeq 8.5$ kpc,
scale height $H \simeq 0.3$ kpc, scale length $h \simeq 3.5$ kpc
and local mass density $\rho_D(R_0) \simeq  1.67 \times 10^8~M_{\odot}$
kpc$^{-3}$.

The dark halo in our Galaxy is also assumed to follow an
isothermal profile
with core radius $a \simeq 5.6$ kpc and local dark matter density
$\rho_H(R_0) \simeq 1.09 \times 10^7~M_{\odot}$ kpc$^{-3}$.
The corresponding asymptotic rotational velocity
is $v_{rot} \simeq 220$ km s$^{-1}$. The MW halo is truncated
at $R \simeq 100 $ kpc.

For both M31 and MW halos, the fraction of dark matter
in form of MACHOs is assumed to be $f_{\rm h}
\simeq 0.2$ \citep{Alcock00}. However, most of our results
can easily be rescaled to get the corresponding figures
for other values of $f_{\rm MACHO}$.

Moreover, we assume that the random velocities of stars and MACHOs
follow Maxwellian distributions
with one-dimensional velocity dispersion
$\sigma = 30, 100, 166$ km s$^{-1}$
and $30, 156$ km s$^{-1}$
for the M31 disk, bulge, halo and MW disk and halo, respectively.
In addition, a M31 bulge rotational velocity of 30 km s$^{-1}$
has been taken into account \citep{Kerins01,An}.

\subsection{Stellar luminosity function}

Pixel lensing event detection by the MEGA collaboration
is performed in the red band
\footnote{Indeed, observations in the red band, by minimizing
light absorption in M31 and MW disks by the intervening dust,
offer the best compromise between sampling and sky background.
Observations in other bands (B and V) are commonly used
to test achromaticity of the candidate events.}
and, thus, red giants are the most luminous stars
in this band. Therefore, we may safely assume that the overwhelming majority
of the pixel lensing event sources are red giants.

Moreover, in the lack of precise information about
the stellar luminosity function in M31, we adopt the
luminosity function derived from the stars in the Galaxy
and assume that it also holds for M31.

Accordingly, following \citep{MamonSoneira} we assume that
the stellar luminosity function does not depend on the position and
in the magnitude range $-6 \le M \le 16$
it is proportional to the expression
\begin{equation}
\phi_*(M) \propto
\frac{ 10^{\beta(M-M^*)}  }
               { [ 1+10^{-(\alpha-\beta)\delta(M-M^*)}]^{1/\delta} }~,
\label{slf}
\end{equation}
where, in the red band,
$M^*= 1.4$, $\alpha \simeq 0.74$, $\beta = 0.045$ and $\delta= 1/3$.
Moreover, the fraction of red giants (over the total star number)
as a function of $M$
is approximated as \citep{MamonSoneira}
\begin{eqnarray}
f_{RG}(M)
&=& 1 - C \exp[\alpha(M+\beta)^{\gamma}]~~~{\rm for}~~-6 \le M \le 3
\nonumber\\
     &=&  0~~~~~~~~~~~~~~~~~~~~~~~~~~~~~~{\rm for}~~M \ge 3~,
\label{fragi}
\end{eqnarray}
where, in the red band,
$C \simeq 0.31$, $\alpha \simeq 6.5 \times 10^{-4}$,
$\beta = 7.5$ and $\gamma \simeq 3.2$.

Therefore, the red giant luminosity function will be
$\phi_{\mathrm RG} (M) \propto \phi_*(M) \times  f_{RG}(M)$ and
the fraction of red giants averaged on the magnitude
is given by
\begin{equation}
\langle f_{RG} \rangle  =
\frac { \int_{-6}^{3} \phi_{\mathrm RG}(M) dM }
{\int_{-6}^{16} \phi_*(M) dM }~\simeq 5.3 \times 10^{-3},
\end{equation}
from which it follows that the local number density of red giants will be
$n_{\mathrm RG} \simeq 5.3 \times 10^{-3}~n_*$, where $n_*$ is the local
stellar number density.

\subsection{Mass functions}

As concerns the lens mass function $\psi_0(\mu)$
in eqs. (\ref{psi0}) - (\ref{gamma11}),
for lenses belonging to the bulge and disk star populations,
the lens mass is assumed to follow a broken power law \citep{GBF}
\begin{eqnarray}
\psi_0(\mu) &=&
K~ \mu^{-0.56}~~~{\rm for}~~0.1 \le \mu \le 0.59
\nonumber\\
&=&
K~ \mu^{-2.20}~~~{\rm for}~~0.59 \le \mu \le
\mu_{\mathrm {up}}
\label{smf}
\end{eqnarray}
where the lower limit $\mu_{\mathrm {min}}=0.1$ and the
upper limit $\mu_{\mathrm {up}}$ is $1$
for M31 bulge stars and $10$ for M31 and MW disk stars
(see also Kerins et al., \citeyear{Kerins01}).
$K$ is fixed according to the normalization given by eq. (\ref{norm}).
The resulting mean mass for lenses in the bulges and disks are
$\langle m_b \rangle \sim 0.37~M_{\odot}$ and
$\langle m_d \rangle \sim 0.69~M_{\odot}$, respectively.

For the lens mass in the M31 and MW halos we
assume the $\delta$-function approximation
\begin{equation}
\psi_0( \mu) = \frac{\rho_0}{M_{\odot}\mu_{\mathrm {h}}}
\delta(\mu-\mu_{\mathrm {h}})
\label{deltamuh}
\end{equation}
and we take
a MACHO mass, in solar units, $\mu_{\mathrm h}= 10^{-2},~10^{-1},~0.5,~1$.

\section{Analytical results}

In the present analysis we adopt the parameters for the INT
and the Sloan-r filter on the WFC (Wide-Field Camera)
used by the MEGA collaboration \citep{Mega04}.
The Telescope diameter, the pixel field of view
and the image exposition time are $2.5$ m, $0.33$ arcsec and $t_{exp}= 760$ s,
respectively.
We also use a gain or conversion factor of 2.8 e$^{-}$/ADU,
and a loss factor $\simeq 3$, for both atmospheric and instrumental.
The zero-point with the Sloan-r WFC turns out to be
$\sim 24.3$ mag arcsec$^{-2}$ \citep{Belokurov}.
Moreover, we adopt a value $\simeq 1.5$ arcsec for the average seeing
conditions, a sky background $ m_{\rm sky} \simeq 20.9$ mag
arcsec$^{-2}$ (corresponding to a Moon eclipse)
and a minimum noise level of
$\sim 2.5 \times 10^{-3} N_{\mathrm {bl}}$.
$N_{\mathrm {bl}}$
is the baseline photon count which is the sum of the M31 surface brightness
given by \cite{Kent} and the sky contribution.

Maps of optical depth, expected event number and event timescale
in pixel lensing experiments have been presented in a previous paper
\citep{dinz} together with the study of the
dependence on microlensing quantities with the assumed M31 mass
distribution model (see also Kerins \citeyear{Kerins04}).
Our results are also in good agreement with previous
analytical estimates for the rate, the timescale distribution
and the optical depth \citep{Baltz02,Gyuk00}.

In Fig. \ref{uth} the map of
$\langle u_{\mathrm T} (x,y) \rangle_{\phi_{\mathrm RG}}$ shows that
in the central M31 regions the lens impact parameter
on average is $\le 0.04$
implying that only high magnified events with $A_{\mathrm T} \ge 25$
are in principle detectable.
The asymmetrical shape is due to the
internal extinction of the M31 disk for which we use the value 0.74 mag
given by \cite{Kent}.
Indeed, due to the inclination of the M31 disk,
along a line of sight towards the southern region there exists
a larger number of source stars
(with respect to the corresponding northern field)
which are not absorbed by the M31 disk dust (and therefore appear
with a smaller magnitude),
lying at larger averaged values for $u_{\mathrm T}$.

%%%%%%%%%%%%%%%%%%%%%%%%%%%%%%%%%%%%%%%%%%%%%%%%%%%%%%%%
\begin{figure}[htbp]
\vspace{9.0cm} \includegraphics{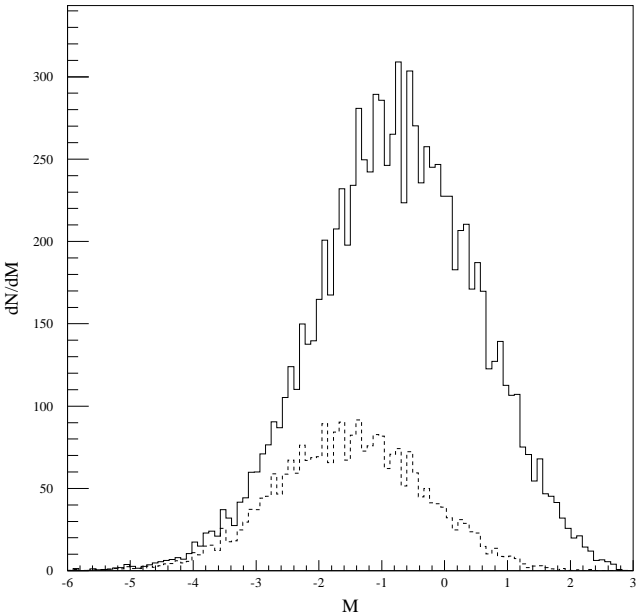}
\caption{Plot of the source absolute magnitude distribution
for the generated events (solid line)
and for the revealed events (dashed line)
towards the MEGA 7 direction.}
\label{mag}
\end{figure}
%%%%%%%%%%%%%%%%%%%%%%%%%%%%%%%%%%%%%%%%%%%%%%%%%%%%%%%%

In Fig. \ref{prodar}, assuming
a MACHO mass $\mu_{\mathrm h}= 0.5$ and a halo MACHO fraction
$f_{\rm h} =0.2$,  we show the map
of the MACHO-to-self lensing probability
ratio $(P_{\mathrm h}/P_{\mathrm s})_{\mathrm {An}}$.
We find that in the M31 central regions microlensing is dominated by
self-lensing contributions while MACHO-lensing becomes important
at distances $>10$ arcmin from the center.
In the figure it is also evident the well known
near-far disk asymmetry due to the inclination of the M31 disk
\citep{Crotts92,Baillon93,Jetzer94}.

%%%%%%%%%%%%%%%%%%%%%%%%%%%%%%%%%%%
\begin{table}
%\centering
\caption{MACHO-to-self lensing probability ratio
$(P_{\mathrm h}/P_{\mathrm s})_{\mathrm {An}}$
towards the 14 MEGA events are given for
different MACHO mass values.}
\medskip
%\begin{center}
\begin{tabular}{|c|c|c|c|c|}
\hline
MEGA & $ \mu_{\mathrm h} =0.01 $ & $ 0.1$ & $0.5$ & $1$ \\
\hline
    1     &    1.65    &    0.52    &    0.23    &    0.16 \\
    2     &    1.76    &    0.56    &    0.25    &    0.18\\
    3     &    2.22    &    0.71    &    0.31    &    0.22\\
    4     &    2.52    &    0.78    &    0.35    &    0.24\\
    5     &   24.58    &    7.72    &    3.46    &    2.40\\
    6     &   23.46    &    7.53    &    3.38    &    2.40\\
    7     &   12.93    &    3.83    &    1.80    &    1.26\\
    8     &    9.79    &    3.15    &    1.40    &    0.96\\
    9     &   19.49    &    6.15    &    2.79    &    1.99\\
   10     &   17.19    &    5.44    &    2.42    &    1.73\\
   11     &   19.22    &    6.04    &    2.74    &    1.88\\
   12     &   18.91    &    5.88    &    2.62    &    1.91\\
   13     &   54.13    &   17.60    &    7.87    &    5.67\\
   14     &  112.32    &   35.47    &   17.07    &   11.12\\
\hline
\end{tabular}
%\end{center}
\label{probrate}
\end{table}

%%%%%%%%%%%%%%%%%%%%%%%%%%%%%%

In Tab. \ref{probrate}, we show how the MACHO-to-self lensing probability
ratio $(P_{\mathrm h}/P_{\mathrm s})_{\mathrm {An}}$
depends on the MACHO mass $\mu_{\mathrm h}$ towards the 14 MEGA events.
The general trend is that
$(P_{\mathrm h}/P_{\mathrm s})_{\mathrm {An}}$
increases as $\mu_{\mathrm h}$ decreases as a consequence of the increase
of the MACHO number density. Moreover, for a given value of $\mu_{\mathrm h}$,
$(P_{\mathrm h}/P_{\mathrm s})_{\mathrm {An}}$
increases with the distance from the M31 center.

%%%%%%%%%%%%%%%%%%%%%%%%%%%%%%%%%%%
\begin{table}
%\centering
\caption{MACHO-to-self lensing Einstein time ratio
$\langle t_{E~\mathrm h} \rangle / \langle t_{E~\mathrm s} \rangle$
towards the 14 MEGA events
are given for different MACHO mass values.}
\medskip
%\begin{center}
\begin{tabular}{|c|c|c|c|c|}
\hline
MEGA & $\mu_{\mathrm h}=0.01 $ & $0.1$ & $0.5$ & $1$ \\
\hline
    1     &    0.30    &    0.96    &    2.15    &    3.02 \\
    2     &    0.31    &    1.00    &    2.42    &    3.24\\
    3     &    0.29    &    0.94    &    2.02    &    2.91\\
    4     &    0.20    &    0.65    &    1.47    &    2.12\\
    5     &    0.16    &    0.51    &    1.10    &    1.59\\
    6     &    0.16    &    0.49    &    1.09    &    1.55\\
    7     &    0.15    &    0.52    &    1.10    &    1.58\\
    8     &    0.13    &    0.41    &    0.92    &    1.35\\
    9     &    0.13    &    0.42    &    0.97    &    1.34\\
   10     &    0.17    &    0.54    &    1.21    &    1.69\\
   11     &    0.16    &    0.50    &    1.11    &    1.61\\
   12     &    0.14    &    0.44    &    0.98    &    1.36\\
   13     &    0.16    &    0.49    &    1.08    &    1.51\\
   14     &    0.15    &    0.48    &    1.03    &    1.52\\
\hline
\end{tabular}
%\end{center}
\label{dtoste}
\end{table}

%%%%%%%%%%%%%%%%%%%%%%%%%%%%%%

In Fig. \ref{timeratio} the map of the MACHO-to-self lensing Einstein
time ratio $(t_{E~\mathrm h}/t_{E~\mathrm s})_{\mathrm {An}}$
is given for a MACHO mass of $0.5~M_{\odot}$.
In the region inside $\simeq $ 5 arcmin
from the M31 center
microlensing events due to MACHOs have twice as long
a duration as compared
to self-lensing events, while in the regions far away from the
M31 center all events have roughly the same duration.
Obviously, the  MACHO-to-self lensing
Einstein time ratio depends on the assumed
MACHO mass. This fact is clearly seen in Tab. \ref{dtoste}, where
the ratio $(<t_{E~\mathrm h}>/<t_{E~\mathrm s}>)_{\mathrm {An}}$
towards the 14 MEGA events is given for different MACHO mass values.
$\mu_{\mathrm h}=0.5$ corresponds to the MACHO mass value for which
the 10 outer MEGA events (5-14)
are characterized by having a MACHO-to-self lensing
Einstein time ratio equal to unity, i.e.,
events due to either MACHO-lensing or
self-lensing have roughly the same duration.
Thus, these different types of
events are indistinguishable on the basis of
their timescale alone.
The situation is instead much more favorable for MACHO masses smaller
or larger than 0.5 $M_{\odot}$.

We would like to emphasize that in pixel lensing experiments
$t_{E}$ is not a directly observable quantity, since
the relevant time scale is
the full-width half-maximum event duration
$t_{1/2}$,
which depends on $t_E$ and the impact parameter $u_0$ \citep{Gondolo}.
However, since the probability for a given $u_0$ value
is practically the same for self-lensing and MACHO-lensing events
(as we have verified by using the MC code)
the ratio  $(<t_{E~\mathrm h}>/<t_{E~\mathrm s}>)_{\mathrm {An}}$
is equivalent
to the ratio  $(<t_{1/2~\mathrm h}>/<t_{1/2~\mathrm s}>)_{\mathrm {An}}$,
at least for the MC generated events.
Of course, as it will be more clear in the following Sections
(in particular from Tab. \ref{tabt12mc} and Fig. \ref{fig10}),
it does not mean that the same conclusion holds for the MC revealed events,
since the (normalized) event number which pass the MEGA selection criteria
turns out to depend on  $t_{1/2}$ and $R_{\mathrm {max}}$
(see Fig. \ref{ed}).

\section{Monte Carlo simulation}

%\subsection{Event generation}

Once the event location (one of the 14 MEGA events towards M31)
has been selected, for any lens and source population
present along the line of sight
we have to make as next the choice over the following five
parameters: source distance $D_{\mathrm os}$, lens distance $D_{\mathrm ol}$,
lense effective transverse velocity
$v_{\mathrm l \perp}$, lens mass $\mu$ and
source magnitude $M$. We shall denote these parameters by $x_i$, with
$i=1,..,5$
in the order just listed (e.g. $x_1=D_{\mathrm os}$, etc.).
The probability with which we select the events
according to one of the parameters is then given by
\begin{equation}
P(x_i) dx_i=
\frac{1}{N_{\mathrm ev}} \frac{\partial N_{\mathrm ev}}{\partial x_i} dx_i
\end{equation}
where
$\frac{\partial N_{\mathrm ev}}{\partial x_i}$
is defined as being the integrand of the event number
$N_{\mathrm ev}$ integrated over
all variables $x_j$ without, however, the considered variable $x_i$.
Clearly when integrating $\frac{\partial N_{\mathrm ev}}{\partial x_i}$ over
$dx_i$ we obtain again $N_{\mathrm ev}$.

%%%%%%%%%%%%%%%%%%%%%%%%%%%%%%%%%%%%%%%%%%%%%%%%%%%%%%%%
\begin{figure}[htbp]
\vspace{9.0cm} \includegraphics{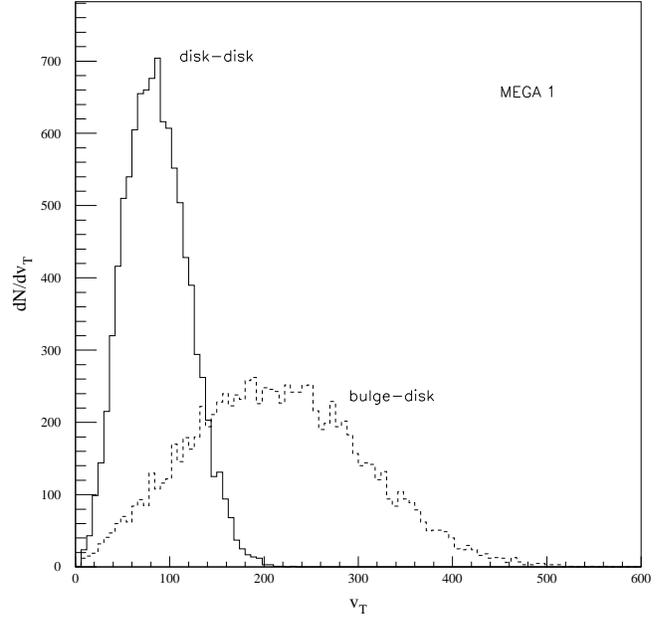}
\caption{The lens transverse velocity $v_{\mathrm l \perp}$
distribution for the MC generated events towards
the MEGA 1 direction is shown
in the case of disk-disk (solid line)
and bulge-disk events (dashed line).}
\label{veltra}
\end{figure}
%%%%%%%%%%%%%%%%%%%%%%%%%%%%%%%%%%%%%%%%%%%%%%%%%%%%%%%%
%%%%%%%%%%%%%%%%%%%%%%%%%%%%%%%%%%%%%%%%%%%%%%%%%%%%%%%%
\begin{figure}[htbp]
\vspace{7.0cm} \includegraphics{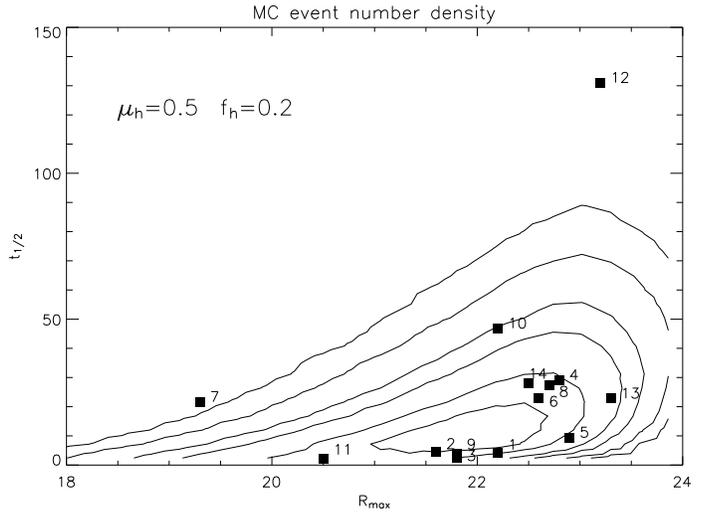}
\caption{
MC revealed event number density $\cal N_{\rm {ev}}^{~\rm {rev}}$
plot (normalized to the maximum value)
as a function of $t_{1/2}$ and $R_{max}$.
Here we consider both self-lensing and MACHO-lensing events,
averaged over all the 14 MEGA directions and we
assume $\mu_{\mathrm h} =0.5$ and $f_{\rm h} =0.2$.
From the inner to the outer
part  $\cal N_{\rm ev}^{~\rm rev}$ decreases with lines referring
to values 0.7, 0.5, 0.3, 0.2, 0.1, 0.05.}
\label{ed}
\end{figure}
%%%%%%%%%%%%%%%%%%%%%%%%%%%%%%%%%%%%%%%%%%%%%%%%%%%%%%%
%%%%%%%%%%%%%%%%%%%%%%%%%%%%%%%%%%%%%%%%%%%%%%%%%%%%%%%%
\begin{figure}[htbp]
\vspace{9.0cm} \includegraphics{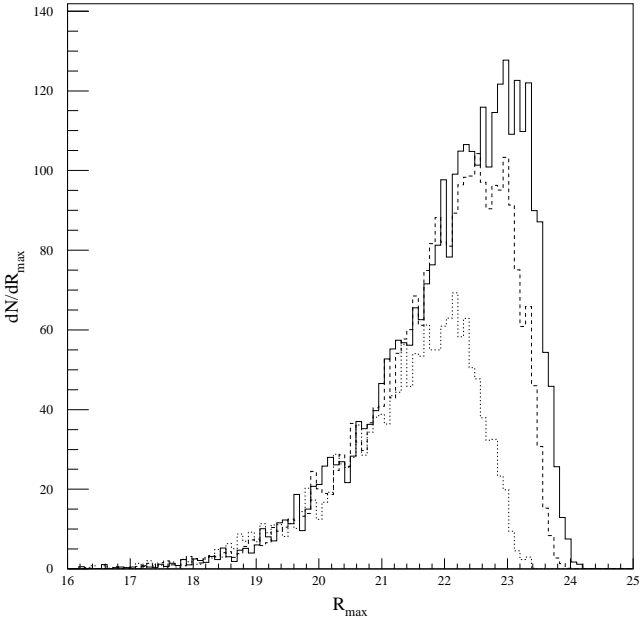}
\caption{The $R_{\mathrm {max}}$
distribution for the MC revealed events towards the MEGA
14 (solid line), 7 (dashed line) and 1 (dotted line) directions.}
\label{rmax}
\end{figure}
%%%%%%%%%%%%%%%%%%%%%%%%%%%%%%%%%%%%%%%%%%%%%%%%%%%%%%

This way, for instance, for self-lensing events
the probability of extracting a lens with mass
$\mu$ turns out to be
\begin{equation}
P(x_4=\mu) d\mu \propto \mu^{1/2} ~ \psi_0(\mu) d\mu~,
\end{equation}
where $\psi_0(\mu)$ is the lens number density distribution defined
in eq. (\ref{smf}).

The probability of a source
\footnote{
Here we take into account that the source number inside the pixel
solid angle increases with the distance as $D_{\mathrm os}^{~~2}$.}
at distance $D_{\mathrm os}$
and a lens at distance $D_{\mathrm ol}$ from the observer,
respectively, is given by

\begin{equation}
P(D_{\mathrm os}) dD_{\mathrm os}
\propto
\rho_{\mathrm s}(D_{\mathrm os})
D_{\mathrm os}^{3/2}
\left(\int_0^{D_{\mathrm os}}
P(D_{\mathrm ol}) dD_{\mathrm ol} \right) dD_{\mathrm os}
\end{equation}
and
\begin{equation}
P(D_{\mathrm ol})
dD_{\mathrm ol}
\propto  \rho_{\mathrm l}
[D_{\mathrm ol} (D_{\mathrm os}-D_{\mathrm ol} )]^{1/2}
(D_{\mathrm ol}) ~ dD_{\mathrm ol}~,
\end{equation}
where $\rho_{\mathrm l}(D_{\mathrm ol})$ and $\rho_{\mathrm s}(D_{\mathrm os})$
are, respectively, the lens and the source
mass densities.

The probability of extracting a source with magnitude $M$ is
weighted by the function
\begin{equation}
P(M) dM\propto u_{\mathrm T} (M) \phi_{\mathrm RG}(M) dM~,
\label{probmag}
\end{equation}
where $\phi_{\mathrm RG} (M)$ is the source magnitude distribution
for red giants defined by means of eqs. (\ref{slf}) - (\ref{fragi})
and, through $u_{\mathrm T}(M)$, we take into account the surface
brightness variation along the field of view.
Indeed, we find that the median value of the source magnitude distribution
decreases from $M^{\mathrm {median}}=-2.18$ for the innermost MEGA directions
to $M^{\mathrm {median}}=-1.18$ for the outermost ones.
In Fig. \ref{mag} we show the distribution of the source absolute magnitude
for all MC generated events and for the revealed (see after) events
towards the direction of the MEGA 7 event.
For a given source and lens position
the probability of extracting a lens with transverse
velocity $v_{\mathrm l \perp}$ is given
by eq. (\ref{provt}), where $z = v_{\mathrm l \perp} / \sigma_{\mathrm l}$.
In Fig. \ref{veltra}, for both disk-disk and bulge-disk events,
we show the lens transverse velocity
distributions for MC generated events towards
the MEGA 1 direction. The shorter $<v_{\mathrm l \perp}>$ value in the first
case will lead to an increase in the event timescale.

The probability of extracting an event with impact parameter
$x_5=u_0$ is constant for values of $u_0$ in the interval
in between 0 and $u_{\mathrm T}$, as defined in eq. (\ref{utmediosuphi})
\footnote
{In order to avoid a sharp cutoff in several obtained plots
(see, e.g., Fig. \ref{rmax}) we allow for events with $u_0 \ut > u_{\mathrm T}$
to be detected.}. Thus it is assumed that
\begin{equation}
P(u_0) \propto {\rm const}.
\end{equation}

Once all the parameters have been fixed, for each event we build the
corresponding lightcurve using the same
time sampling of the MEGA campaign
and the same observing and instrumental conditions of the considered
experiment.
We add a gaussian noise modulated by the Moon phase
and use a Paczy\'nski fit \citep{Paczynski}
to evaluate the microlensing parameters.

The selection of the MC generated events is based on the same criteria
adopted by the MEGA collaboration.
By using the results of the Paczy\'nski fit
we filter the lightcurves
with the following selection criteria:
{\it peak sampling},
{\it peak significance},
{\it peak width},
{\it baseline flatness} and
{\it goodness of fit}
(for more details see Appendix A.1 in \citealt{Mega04}).
The criteria that more severely cut the MC generated events are
the {\it peak sampling} and the {\it peak significance},
which depends through the evaluation of the statistical errors on the
the Moon phase.
The former criterium
leads to select events with well sampled lightcurves
(within the MEGA campaign), the latter to events with
high signal-to-noise ratios.
The fraction of MC selected events also depends on
the event location through the threshold value of $R_{\mathrm {max}}$.
This effect is further discussed in the following section.

%%%%%%%%%%%%%%%%%%%%%%%%%%%%%%%%%%%%%%%%%%%%%%%%%%%%%%%%
\begin{figure}[htbp]
\vspace{9.0cm} $\begin{array}{c}
\includegraphics{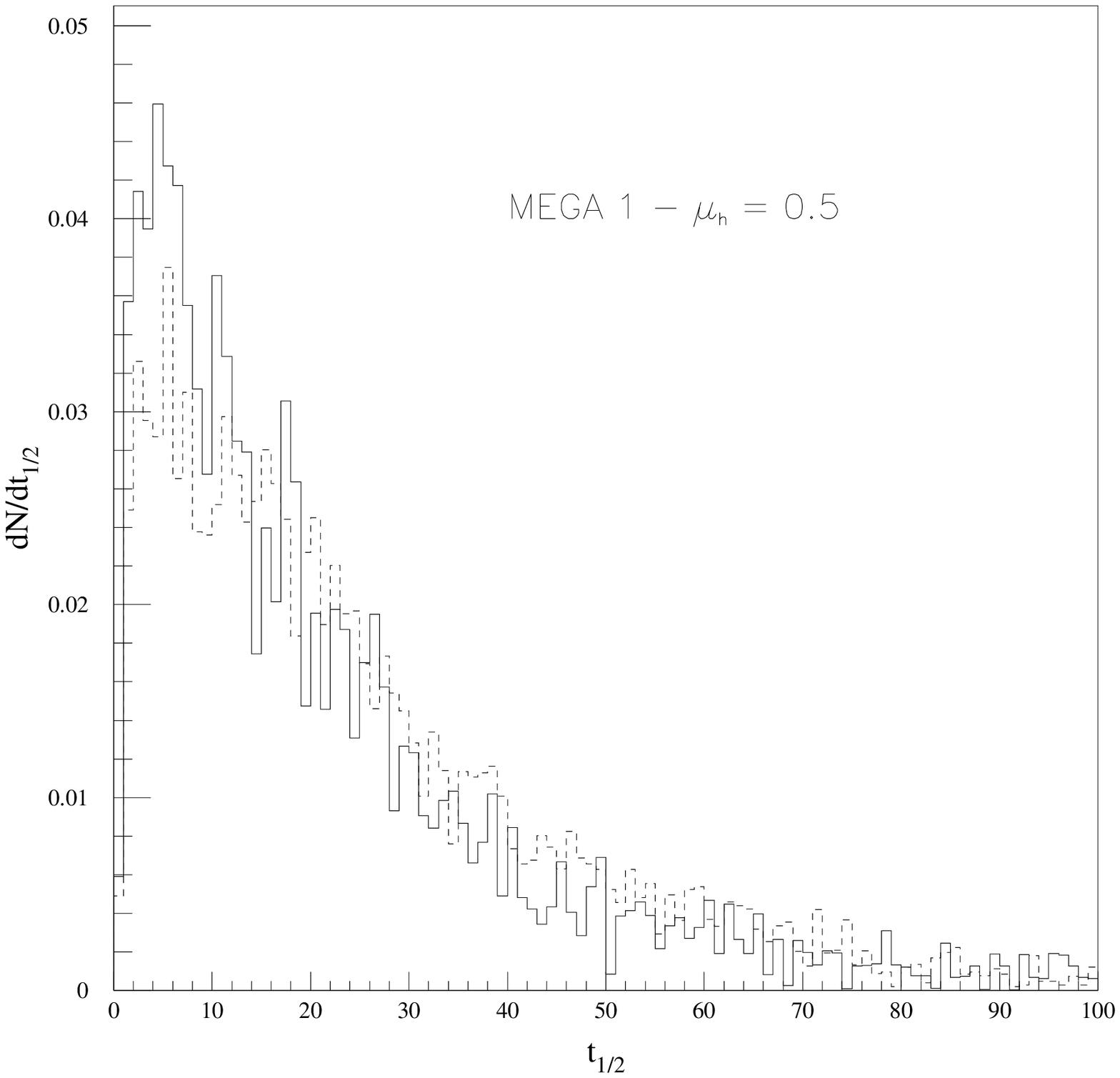}
\\[9.0cm]
\includegraphics{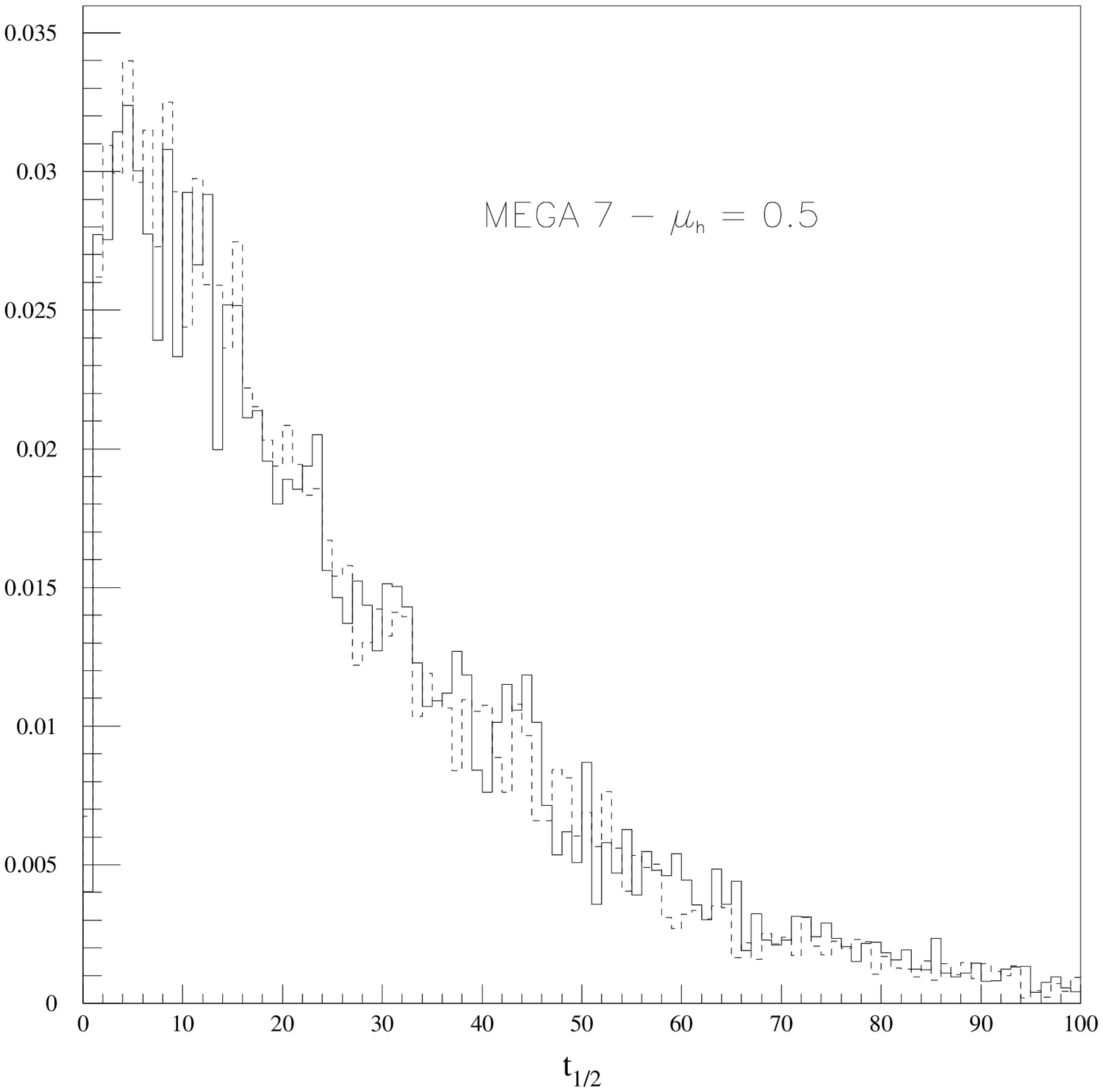}
\end{array}$
\caption{The $t_{1/2}$ distributions are shown for MEGA 1 and 7 events
and for $\mu_{\mathrm h} = 0.5$.
In each panel, the solid curve refers to self-lensing events and
the dashed curve to MACHO-lensing events.}
\label{t12}
\end{figure}
%%%%%%%%%%%%%%%%%%%%%%%%%%%%%%%%%%%%%%%%%%%%%%%%%%%%%%%%

\section{Monte Carlo results}

MC results allow to estimate the features of the revealed
events which pass the adopted selection criteria.
In Fig. \ref{ed}, assuming a MACHO mass $\mu_{\mathrm h}= 0.5$ and a
halo MACHO fraction $f_{\rm h} =0.2$ we give
the contour plot, in the $(t_{1/2},R_{\mathrm {max}})$ parameter space,
of the event number density
\begin{equation}
{\cal N}_{\rm {ev}}^{~\rm {rev}} =
\frac{ d^2 ~ N_{\mathrm {ev}}^{~\mathrm {rev}}}
{d t_{1/2} dR_{\mathrm{max}}}~,
\end{equation}
averaged on all the 14 MEGA directions,
for both self-lensing and MACHO-lensing  events.
We also give in the same parameter space the position of the 14 observed
MEGA events.
$\cal N_{\rm {ev}}^{~\rm {rev}}$ is maximum in the region
$t_{1/2} \simeq 20$ day and $R_{\mathrm {max}} \simeq 22$.
Moreover, $\cal N_{\rm {ev}}^{~\rm {rev}}$ rapidly decreases
for $R_{\mathrm {max}} > 23.5$ (due to the lack of revealed events with low
signal-to-noise ratio) and $t_{1/2}<5$ day (due to the peak sampling)
and in the region of high amplification and long duration events (upper-left
region of the figure), due to the absence of MC generated events.
Actually, the last cutoff depends on the adopted MACHO mass value
and shifts towards smaller $t_{1/2}$ values decreasing $\mu_{\mathrm h}$.
Indeed, in Fig. \ref{ed} the crucial parameter
determining the event distribution
is the lens mass value and it turns out that
the region where $\cal N_{\rm {ev}}^{~\rm {rev}}$
is maximum scales with $\mu_{\mathrm h}$ in the same way as
$t_{1/2}^{~~~\mathrm {median}}$ and
$R_{\mathrm {max}}^{~~~\mathrm {median}}$
(see Tab. \ref{tabt12mc}, where we give the median values of
the distributions of the MC revealed events
as a function of the lens star mass $\mu_*$,
$t_{1/2}$ and $R_{\mathrm {max}}$).
%%%%%%%%%%%%%%%%%%%%%%%%%%%%%%%%%%%%%%%%%%%%%%%%%%%%%%%%%%%%
\begin{figure}[htbp]
\vspace{9.0cm} $\begin{array}{c}
\includegraphics{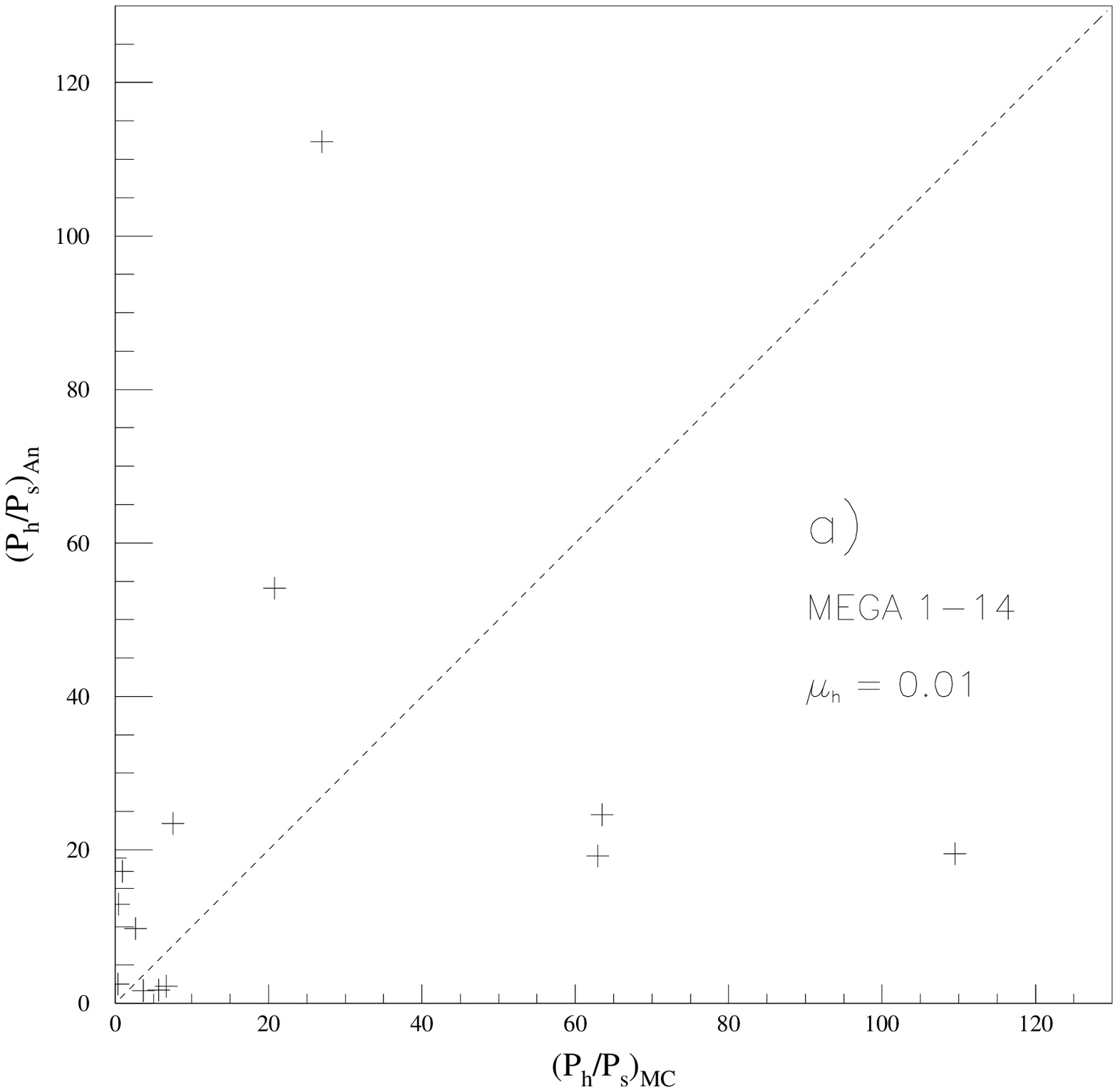}
\\[9.0cm]
\includegraphics{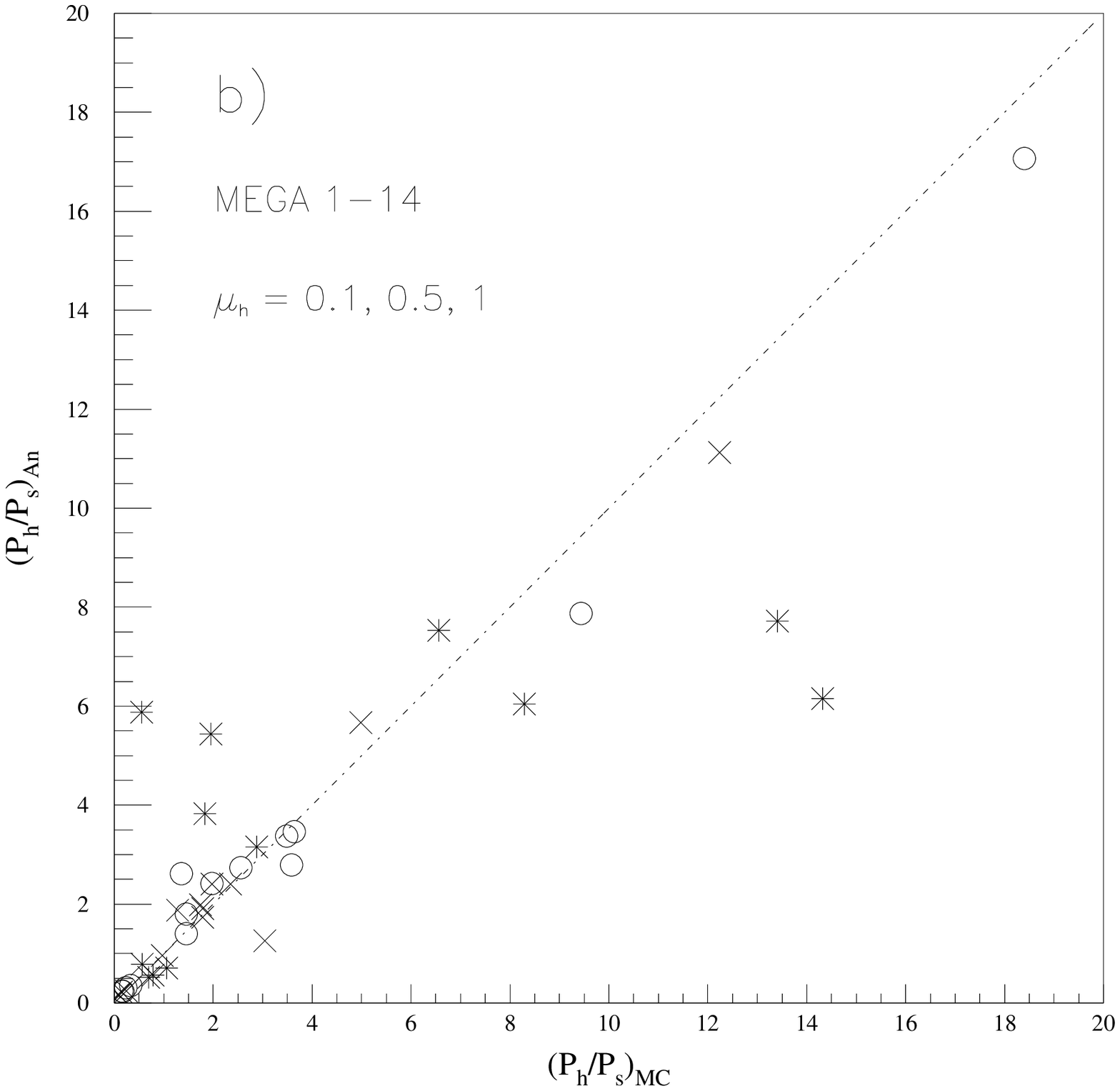}
\end{array}$
\caption{Plot of $(P_{\mathrm h}/P_{\mathrm s})_{\mathrm {An}}$ versus
$(P_{\mathrm h}/P_{\mathrm s})_{\mathrm {MC}}$ for all the
14 MEGA events.
In panel a) we consider a MACHO mass $\mu_{\mathrm h}=0.01$ and
we use the symbol plus; in panel b) we take $\mu_{\mathrm h}=0.1,~ 0.5,~ 1$
and use symbols time, circle, asterisk, respectively.}
\label{fig9}
\end{figure}
%%%%%%%%%%%%%%%%%%%%%%%%%%%%%%%%%
%%%%%%%%%%%%%%%%%%%%%%%%%%%%%%%%%%%%%%%%%%%%%%%%%%%%%%%%%%%%
\begin{figure}[htbp]
\vspace{9.0cm} $\begin{array}{c}
\includegraphics{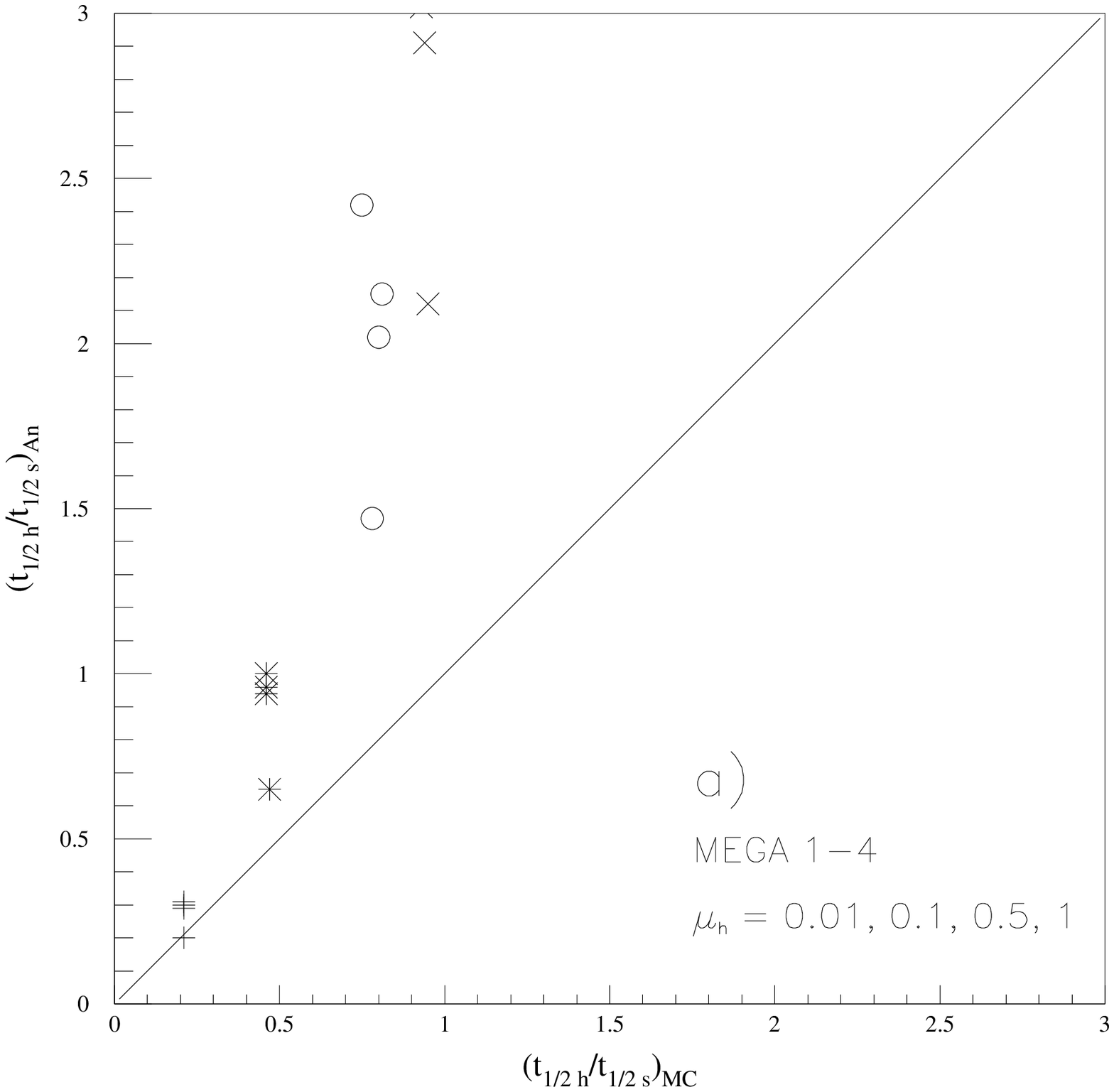}
\\[9.0cm]
\includegraphics{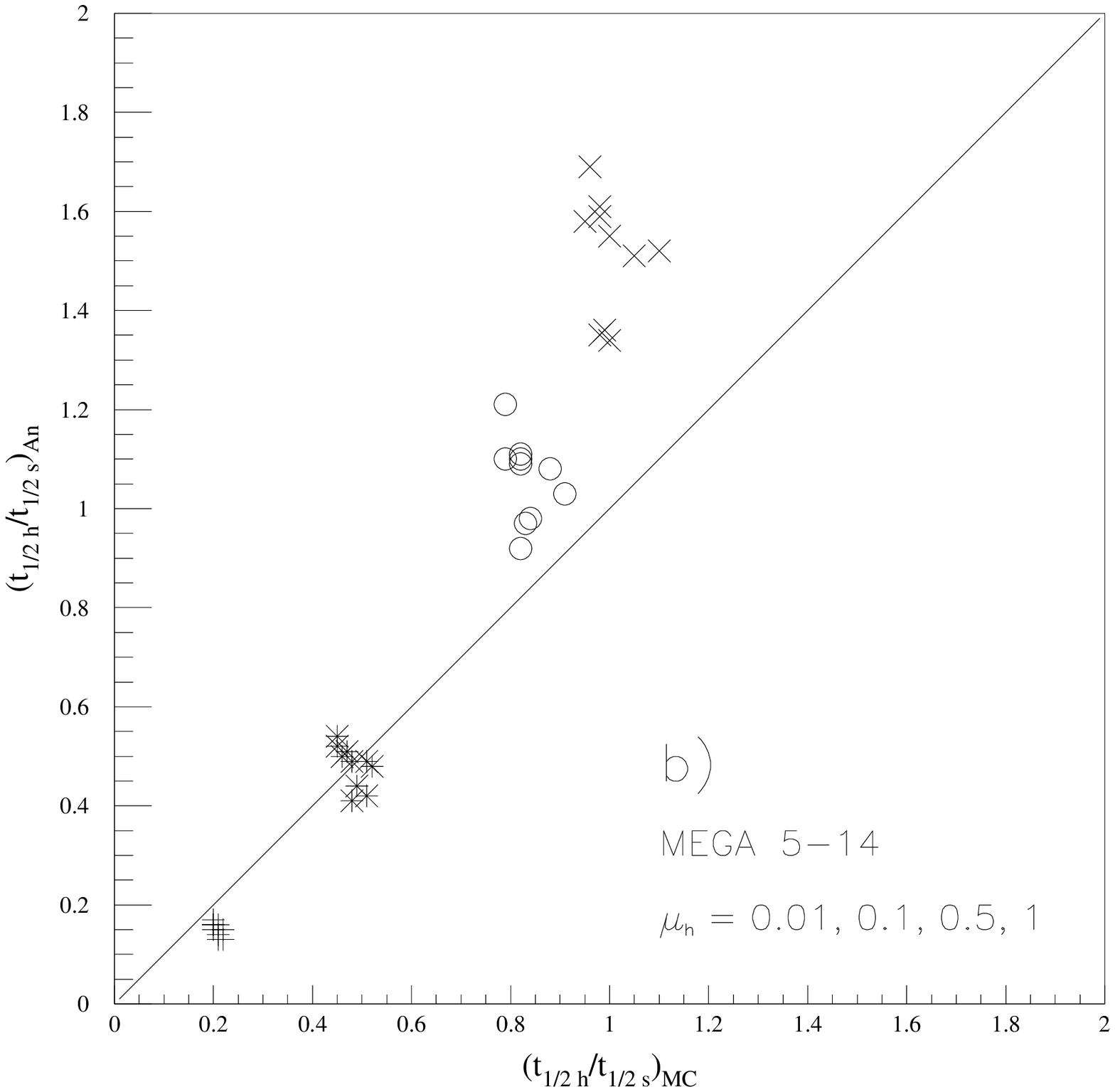}
\end{array}$
\caption{
Plot of $(t_{1/2~\mathrm h}/t_{1/2~\mathrm s})_{\mathrm {An}}$ versus
$(t_{1/2~\mathrm h}/t_{1/1~\mathrm s})_{\mathrm {MC}}$, for
$\mu_{\mathrm h}=0.01, 0.1, 0.5, 1$. The symbols are as in Fig. \ref{fig9}.
In panel a) we consider MEGA 1-4 and
in panel b) MEGA 5-14 events, respectively.}
\label{fig10}
\end{figure}
%%%%%%%%%%%%%%%%%%%%%%%%%%%%%%%%%

%%%%%%%%%%%%%%%%%%%%%%%%%%%%%%%%%%%%%%%%%%%%%%%%%%%%%%%
\begin{table*}
%\centering
\caption{
For the MC revealed events (which pass the MEGA selection criteria),
the median values
$\mu_*^{~\mathrm {median}}$ (column 2)
of the lens star mass distribution
for self-lensing events,
$t_{1/2}^{~~~\mathrm {median}}$ (columns 3-7) and
$R_{\mathrm {max}}^{~~~\mathrm {median}}$ (columns 8-12)
for self-lensing and MACHO-lensing
are given towards the 14 MEGA events.}
\medskip
%\begin{center}
\begin{tabular}{|c||c||ccccc||ccccc|}
\hline
         & $\mu_*^{~\mathrm {median}}$&       &     & $t_{1/2}^{~~~\mathrm {median}}$&&       &      &       &
$R_{\mathrm {max}}^{~~~\mathrm{median}}$ & & \\
\hline
         &  self &  self &     &           MACHO &        &        &self   &      & MACHO &      &      \\
MEGA     &       &       & $\mu_{\mathrm h}=0.01$& $ 0.1$ & $ 0.5$ &  $ 1$ &&$\mu_{\mathrm h}=0.01$& $0.1$&$
 0.5$ & $ 1$ \\
\hline
1& 0.48  & 15.66 & 4.96& 11.22&   19.51&   22.49&  21.56& 21.16& 21.45& 21.65& 21.61\\
2& 0.48  & 15.33 & 5.06& 11.22&   18.15&   22.04&  21.62& 21.31& 21.52& 21.64& 21.70\\
3& 0.49  & 16.38 & 5.17& 11.20&   19.32&   22.62&  21.69& 21.41& 21.64& 21.76& 21.74\\
4& 0.62  & 18.16 & 5.04& 11.43&   18.81&   23.04&  21.87& 21.57& 21.85& 21.93& 21.94\\
5& 0.46  & 20.62 & 4.85& 11.43&   19.80&   23.59&  22.22& 21.85& 22.11& 22.23& 22.22\\
6& 0.46  & 21.43 & 4.77& 11.57&   19.88&   24.27&  22.17& 21.85& 22.11& 22.23& 22.23\\
7& 0.50  & 20.72 & 4.95& 10.97&   19.12&   22.96&  22.10& 21.79& 22.02& 22.14& 22.19\\
8& 0.65  & 21.47 & 5.15& 11.60&   19.90&   23.65&  22.13& 21.81& 22.05& 22.18& 22.21\\
9& 0.57  & 24.04 & 5.20& 12.21&   19.94&   24.17&  22.19& 21.85& 22.08& 22.18& 22.20\\
10&0.45  & 20.07 & 4.73& 10.79&   18.98&   23.09&  22.19& 21.83& 22.06& 22.20& 22.24\\
11&0.46  & 20.37 & 4.88& 11.02&   19.79&   23.66&  22.14& 21.86& 22.08& 22.21& 22.22\\
12&0.52  & 22.30 & 5.00& 11.77&   20.24&   23.94&  22.17& 21.84& 22.05& 22.20& 22.21\\
13&0.44  & 22.58 & 4.98& 12.24&   21.14&   25.28&  22.32& 21.99& 22.20& 22.34& 22.35\\
14&0.43  & 24.15 & 5.34& 12.58&   21.90&   26.53&  22.34& 22.03& 22.25& 22.35& 22.38\\
\hline
\end{tabular}
%\end{center}
\label{tabt12mc}
\end{table*}
%%%%%%%%%%%%%%%%%%%%%%%%%%%%%%%%%%%%%%%%%%%%
From Tab. \ref{tabt12mc} for $\mu_{\mathrm h} =0.5$ one can infer that the
$\cal N_{\rm ev}^{~\rm rev}$ plot does not vary substantially
for MACHO-lensing and self-lensing events occurring away from the
M31 center, since the lens stellar mass
is of the same order of the MACHO mass. Moreover, for any value of
$\mu_{\mathrm h}$, the $\cal N_{\rm ev}^{~\rm rev}$ distribution is also
weakly dependent on the selected direction towards M31.
The situation is completely different
for $\mu_{\mathrm h} \ne 0.5$, and this means that
self-lensing and MACHO-lensing events lie on different
regions in the $(t_{1/2},R_{\mathrm {max}})$ parameter space of the
corresponding $\cal N_{\rm {ev}}^{~\rm {rev}}$ plot.

%%%%%%%%%%%%%%%%%%%%%%%%%%%%%%
\begin{table}
%\centering
\caption{MACHO-to-self lensing probability ratios
$(P_{\mathrm h}/P_{\mathrm s})_{\mathrm {MC}}$
towards the 14 observed MEGA events are given for
different MACHO mass values. Probabilities are now calculated
from eq. (\ref{32})
by considering microlensing rates,
$t_{1/2}$ and $R_{\mathrm {max}}$ distributions
for the MC revealed events.
The results in the table scale with the
MACHO fraction value as $f_{\mathrm h}/0.2$.}
\medskip
%\begin{center}
\begin{tabular}{|c|c|c|c|c|c|c|}
\hline

MEGA & $\mu_{\mathrm h}=0.01 $ & $0.1$ & $0.5$ & $1$ \\
\hline
  1&    3.65&    0.70&    0.18&    0.10 \\
  2&    5.71&    0.79&    0.16&    0.08\\
  3&    6.70&    1.06&    0.23&    0.14\\
  4&    0.36&    0.56&    0.33&    0.24\\
  5&   63.51&   13.41&    3.64&    1.98\\
  6&    7.56&    6.56&    3.48&    2.35\\
  7&    0.45&    1.83&    1.46&    3.04\\
  8&    2.67&    2.88&    1.46&    0.97\\
  9&  109.47&   14.32&    3.58&    1.74\\
 10&    0.98&    1.95&    1.98&    1.79\\
 11&   62.95&    8.29&    2.56&    1.28\\
 12&    0.04&    0.55&    1.36&    1.79\\
 13&   20.81&   23.20&    9.44&    4.98\\
 14&   26.99&   29.88&   18.50&   12.24\\
\hline
\end{tabular}
\label{probratemc}
\end{table}

The dependence of $t_{1/2}^{~~~\mathrm {median}}$ and
$R_{\mathrm {max}}^{~~~\mathrm {median}}$
on the lens mass and the MEGA direction is also
clear from Tab. \ref{tabt12mc}.
For self-lensing events, $t_{1/2}^{~~~\mathrm {median}}$ increases from
the inner to the outer part of the M31 galaxy, and in this region
$t_{1/2}^{~~~\mathrm {median}}$ also increases for increasing values of
$\mu_*^{~\mathrm {median}}$, following the dependence of the Einstein radius
with the lens mass
\footnote{For self-lensing events, the lens median mass value
changes with the MEGA direction, since the disk lens mass is on average higher
than the bulge lens mass and the disk-to-bulge probability ratio depends
on the MEGA direction.}.
Moreover, for MACHO-lensing events with a fixed $\mu_{\mathrm h}$ value,
$t_{1/2}^{~~~\mathrm {median}}$ is almost the same for any MEGA direction,
while it increases, as expected, with the lens mass.

In Tab. \ref{tabt12mc} is also evident
the decrease, moving towards the M31 center, of the
$R_{\mathrm {max}}^{~~~\mathrm {median}}$ value
that follows the surface brightness
variation along the field. In particular, we find a shift of about 1 mag
going from the inner to the outer regions.
In Fig. \ref{rmax} we give the (revealed) event distributions as a function
of $R_{\mathrm {max}}$ towards the MEGA 1, 7 and 14 directions.
This clearly shows how the event location affects the fraction of
the revealed events.

The event distributions as a function of $t_{1/2}$,
for both self-lensing (solid line) and
MACHO-lensing (dashed line) revealed events are shown in Fig. \ref{t12},
for $\mu_{\mathrm h} = 0.5$ and
for the 1 and 7 MEGA directions towards M31
(representative of inner and outer events, respectively).
Only in the case of the MEGA 1
direction the obtained  distributions are markedly different
for self-lensing and MACHO-lensing, since self-lensing events are shorter
than MACHO-lensing events.
For the MEGA 7 direction (and the other outer directions)
self-lensing and MACHO lensing with the same lens mean mass
have roughly the same $t_{1/2}$ distributions.

Our MC results can be used to estimate the lens nature and location
of the 14 MEGA candidate events, by
weighting  the microlensing rate -
giving the analytical estimates of the
MACHO-to-self lensing  probability ratio
$(P_{\mathrm h}/P_{\mathrm s})_{\mathrm {An}}$
shown in Tab. \ref{probrate}  -
with the revealed event number density distribution
and taking also into account
the observed features of the MEGA events.
In Tab. \ref{probratemc},
assuming a halo MACHO fraction $f_{\mathrm h} = 0.2$
and different MACHO mass values,
we give the MC MACHO-to-self lensing probability ratio
\begin{equation}
\left(\frac {P_{\mathrm h}}{P_{\mathrm s}}  \right)_{\mathrm {MC}} =
\left(\frac {P_{\mathrm h}}{P_{\mathrm s}}  \right)_{\mathrm {An}} ~
\frac{{\cal N}_{\mathrm {ev~h}}^{~\mathrm {rev}}(t_{1/2}^{~~\mathrm {obs}})}
     {{\cal N}_{\mathrm {ev~s}}^{~\mathrm {rev}}(t_{1/2}^{~~\mathrm {obs}})}
\frac{{\cal N}_{\mathrm {ev~h}}^{~\mathrm {rev}}
(R_{\mathrm {max}}^{~~\mathrm {obs}})}
     {{\cal N}_{\mathrm {ev~s}}^{~\mathrm {rev}}
(R_{\mathrm {max}}^{~~\mathrm {obs}})}
\label{32}
\end{equation}
where we indicate with
${\cal N}_{\mathrm {ev}}^{~\mathrm {rev}} (t_{1/2}^{~~\mathrm {obs}})$
and
${\cal N}_{\mathrm {ev}}^{~\mathrm {rev}}(R_{\mathrm {max}}^{~~\mathrm {obs}})$
the number of events (either MACHO or self-lensing)
with duration and magnitude at maximum, respectively,
within 2 standard deviations around the
observed values.
Here, we remark that each
${\cal N}_{\mathrm {ev}}$ distribution is normalized
to the total number of revealed events.
A comparison between the analytical and MC results given, respectively,
in Tabs. \ref{probrate}
and \ref{probratemc} is presented in Figs. \ref{fig9}a and \ref{fig9}b, where
we plot  $(P_{\mathrm h}/P_{\mathrm s})_{\mathrm {An}}$ versus
$(P_{\mathrm h}/P_{\mathrm s})_{\mathrm {MC}}$,
for the MACHO mass values $\mu_{\mathrm h}=0.01$ and $\mu_{\mathrm h}=0.1,~0.5,1~$, respectively.
Only for $\mu_{\mathrm h}=0.5$ and for the outer events
MC and analytical estimates are in good agreement,
since, as we have already discussed above,
MACHO-lensing and self-lensing events have for this
particular mass value, on average, the same features.

Also for the time scale ratios between MACHO-lensing and self-lensing
events analytical and MC results give different estimates.
A comparison of the analytical results
in Tab. \ref{dtoste} and
MC results in Tab. \ref{tabt12mc}
is presented in Figs. \ref{fig10}a and \ref{fig10}b, where
we show  $(t_{1/2~\mathrm h}/t_{1/2~\mathrm s})_{\mathrm {An}}$ versus
$(t_{1/2~\mathrm h}/t_{1/1~\mathrm s})_{\mathrm {MC}}$
for MEGA 1-4 and 5-14 events.
The difference is particularly important for the events 1-4.

It emerges, therefore, clearly that the MC analysis is essential
for determining the lens nature and location of microlensing
events, at least if the MACHO mass differs substantially
from the average self-lensing mass.
From Tab. \ref{probratemc}, we find that
MEGA events 1-4 are most likely self-lensing events for
MACHO masses greater than 0.1 solar masses,
while events 5, 6, 9, 11, 13, 14 are likely MACHO-lensing.
For the other MEGA directions
the lens nature is more uncertain.
As a final comment on Fig. \ref{ed}, we note that the event 12, lying in a
region of low (revealed)
event number density, hardly can be considered a reliable
microlensing event, unless the MACHO mass is considerably greater than 1
$M_{\odot}$.

\section{Conclusions}

We have studied the main features of the expected microlensing events
in pixel lensing observations towards M31, by using both analytical estimates
(from the microlensing rate) as well as results by a MC code
where we reproduce the observing and instrumental conditions
of the MEGA experiment.

First of all, we derive in Section 2 the microlensing rate, and
assuming a specific mass distribution model for M31 and the Galaxy,
we calculate the MACHO-to-self lensing probability and
the MACHO-to-self lensing event time scale ratios.
For $\mu_{\mathrm h} >0.1$, we find that self-lensing dominates in the
M31 central regions. Moreover, for $\mu_{\mathrm h}  \simeq 0.5$, towards the innermost
MEGA directions, MACHOs events have duration twice as long
as
self-lensing events, while outer events have roughly the same duration.

We then generate a large number of MC microlensing
events by choosing relevant source and lens parameters
as outlined in Section 5.
We study the observability of the MC events,
by considering the capabilities of the INT Telescope,
typical observing conditions and
the event selection criteria adopted by the MEGA Collaboration.

MC results can be used to evaluate, for the 14 MEGA candidate events,
the MACHO-to-self lensing probability and the
event time scale ratios $(P_{\mathrm h}/P_{\mathrm s})_{\mathrm {MC}}$ and
$(t_{1/2~\mathrm h}/t_{1/2~\mathrm s})_{\mathrm {MC}}$
(given in  Tabs. 4 and 5), by taking into account not only the analytical
expectations from the microlensing rate (as already done in Tabs. 2 and 3)
but also the features of the MC (revealed) events and the observed values
of $t_{1/2}$ and $R_{\mathrm {max}}$.
MC results and analytical expectations are compared in Figs. 9 and 10,
where one can see that in determining the lens nature and location of the
MEGA candidate events, the MC analysis is particularly important for
$\mu_{\mathrm h} \ne 0.5$.
Accordingly, we find that event 12, lying in a region of low event number
density, hardly can be considered a reliable microlensing event, unless the
MACHO mass is considerably larger than 1 $M_{\odot}$.
Moreover, for a MACHO mass greater than $0.1~M_{\odot}$,
the innermost MEGA events 1, 2, 3, 4  are most likely self-lensing events,
while 5, 6, 9, 11, 13, 14 are MACHO-lensing events. For the other MEGA directions
the lens nature is more uncertain.

\begin{acknowledgements}
SCN is supported by the Swiss National Science Foundation.
\end{acknowledgements}

%----------------------------------------------------------------------

\end{document}